\documentclass[
aps,%
12pt,%
final,%
notitlepage,%
oneside,%
onecolumn,%
nobibnotes,%
nofootinbib,%
superscriptaddress,%
noshowpacs,%
]%
{revtex4}
\usepackage{epsf}
\usepackage{citesort}
\begin{document}

\title{Model of Centauro and strangelet production in heavy ion collisions}

\author{\firstname{A.\,L.\,S.\,}\surname{Angelis}}
\affiliation{Nuclear and Particle Physics Division,
  Physics Department, University of Athens, Athens, Greece}
\author{E.\,G{\l}adysz-Dziadu\'s}
\affiliation{Laboratory of High Energy
 Physics, Institute of Nuclear Physics, Krakow, Poland}
\author{Yu.\,V.\ Kharlov}
\affiliation{Institute for High Energy Physics,
 Protvino, Russia}
\author{V.\,L.\,Korotkikh}
\affiliation{Moscow State University, SINP, Moscow,
 Russia}
\author{G.\,Mavromanolakis}
\affiliation{Nuclear and Particle Physics Division,
 Physics Department, University of Athens, Athens, Greece}
\author{A.\,D.\,Panagiotou}
\affiliation{Nuclear and Particle Physics Division,
 Physics Department, University of Athens, Athens, Greece}
\author{S.\,A.\ Sadovsky}
\affiliation{Institute for High Energy Physics,
 Protvino, Russia}

\begin{abstract}
  We discuss the phenomenological model of Centauro event production
  in relativistic nucleus--nucleus collisions. This model
  makes quantitative predictions for kinematic observables,
  baryon number and mass of the Centauro fireball and its decay products.
  Centauros decay mainly to nucleons, strange hyperons and  possibly
  strangelets. Simulations of Centauro events for the CASTOR detector
  in $Pb+Pb$ collisions at LHC energies are performed.
  The signatures of these events are discussed in detail.
\end{abstract}

\maketitle

\section*{Introduction}

In this paper we present the Monte Carlo generator of Centauro
events~\cite{Lattes1980,Cosmic}, produced in relativistic
nucleus--nucleus collisions based on the phenomenological model
described in~\cite{Pana89,Pana92,Pana94,Ewa2001}.  Originally the
generator has been used to simulate Centauro production in $\sqrt{s} =
5.5~\times A$~TeV $Pb+Pb$ collisions at the LHC and to study the
performance of the CASTOR detector, which was initially under
development for the ALICE experiment~\cite{ALICE-tp}. Later it became
clear that the infrastructure of the experiment CMS~\cite{CMS-tp} is
more suitable for these studies and the decision has been taken to
carry out this experiment in the CMS.

Originally the model of Centauro event production was based on
experimental facts known from the cosmic ray studies. Experimentally
observed characteristics such as multiplicities, transverse momenta,
energy spectra and pseudorapidity distributions of secondary particles
inspired the scenario of the Centauro fireball evolution through which
the thermodynamical parameters and the lifetime of the Centauro
fireball can be calculated. The extrapolation of this model to higher
energies allowed to estimate some observables of Centauro events at
the LHC energy range taking into account the collider
kinematics~\cite{Gla95}.

In the approach adopted here we attempt to predict more precisely
the characteristics of such events.
We present the quantitative description of the original phenomenological
model of Centauro production in nucleus--nucleus collisions
under the assumption of some fundamental characteristics of the
Centauro fireball, leading to more detailed predictions of
observables in such events.
The model is formulated in terms of the impact parameter of the
nucleus--nucleus collision, two thermodynamical parameters
(baryochemical potential and temperature) which are assigned to the
Centauro fireball and the nuclear stopping power. In order to construct
a fully quantitative model we formalize all assumptions of the original
model and introduce some additional ones. The event generator
{\sc CNGEN} calculates the Centauro fireball parameters and produces the
full event configuration. In this manner the model reproduces
all the kinematical parameters of the Centauro events which were
observed in cosmic ray experiments.

In section 1 we give the thermodynamical and kinematical description
of the production and evolution of Centauro-type events in relativistic
nucleus--nucleus collisions and present some of their
characteristics such as their mass, energy and multiplicity
distributions.

In section 2 we give results on the detection capability of such
events with the CASTOR detector. Centauro events are compared with
conventional events produced by the HIJING generator~\cite{Hijing}.
Signatures of Centauro events are discussed.

\section{Physics of Centauro events}

\paragraph*{Centauro fireball evolution.}
The phenomenological description of Centauro events
was introduced in~\cite{Pana89,Pana92,Pana94}. According to this model
Centauro events are produced in the projectile fragmentation region
of a relativistic nuclear collision when the projectile nucleus penetrating
through the target nucleus transforms its kinetic energy to heat and forms
a relatively cool quark matter state with high baryochemical
potential~\cite{Pana92,Pana94}. We refer to this quark matter state
as the primary Centauro fireball. At the first stage of its evolution
it contains $u$ and $d$ quarks and gluons. The high baryochemical potential
inhibits gluons from fragmenting into $u\bar u$ and $d\bar d$ pairs
due to Pauli blocking~\cite{Pana92}. Therefore gluons fragment
into $s\bar s$ pairs and a state of partial chemical equilibrium is
achieved. During thistime $\bar s$ quarks couple with $u$ and $d$
quarks and a number of $K^+$ and $K^0$ are emitted from the primary fireball, 
decreasing the temperature and removing entropy. At the
end of this stage the Centauro fireball has become a slightly strange
quark matter (SQM) fireball with a relatively long lifetime
($\tau \sim 10^{-9} $~sec)~\cite{Olga2000}.
In the case of cosmic ray Centauros this allows the fireball to reach
mountain top altitudes.
At the same time a mechanism of strangeness separation~\cite{Greiner}
can cause the strange quark content of the SQM fireball to accumulate
in one or more smaller regions inside it.
The SQM fireball finally decays explosively into non-strange baryons
and light ($A > 6$) strange quark matter objects (strangelets).

\paragraph*{Baryon number of the Centauro fireball.}
We consider collisions of nuclei with atomic numbers $A_1$ and $A_2$
and charges $Z_1$ and $Z_2$ respectively. The impact parameter $b$ is
roughly restricted by
$$
0 < b < R_1 + R_2,
$$
where $R_i = 1.15A_i^{1/3}$~fm ($i=1,2$) are the radii of the
colliding nuclei. The Centauro fireball is produced in the region
of overlap of the two nuclei. We assume that all nucleons of the
projectile nucleus which fall within this region can interact, and that
it is these nucleons which define the fireball's baryon number.
The baryon number $N_b$ of the fireball can then be estimated from
simple geometrical considerations.
Assuming a uniform distribution of nucleons in a nucleus one can
obtain $N_b$ through the ratio of the volumes of the overlapping
region $V_{\rm ovrlp}$ and the whole projectile nucleus $V_1$:
%
\begin{equation}
N_b = 0.9\, A_1 \frac{V_{\rm ovrlp}}{V_1}. \label{eq:Nb}
\end{equation}
Where the factor 0.9 has been introduced in order to exclude
the contribution to $N_b$ from the boundaries of the overlapping region.

It is natural to assume that projectile and target nuclei are distributed
uniformly in the transverse plane, i.e.\ that $b^2$ is distributed
uniformly. This assumption determines the shape of further
distibutions presented below.
All cosmic ray Centauro events were observed with
hadron multiplicity $N_h > 70$, hence in our model we restrict the Centauro
fireball production to $N_b > 50$.

In our quantitative model we use the assumption that each
nucleus--nucleus collision produces a Centauro fireball characterized
by the same thermodynamical parameters. This assumption is reasonable
when only small variations of the impact parameter are considered.
Central collisions are more likely to produce the Centauro fireball than
peripheral ones because of the larger baryon content.
Therefore the distributions shown in this paper have been calculated
for $Pb+Pb$ collisions with impact parameter $0~<~b~<~5$~fm.

\begin{figure}[ht]
\parbox{0.48\hsize}{\epsfxsize=\hsize \epsfbox{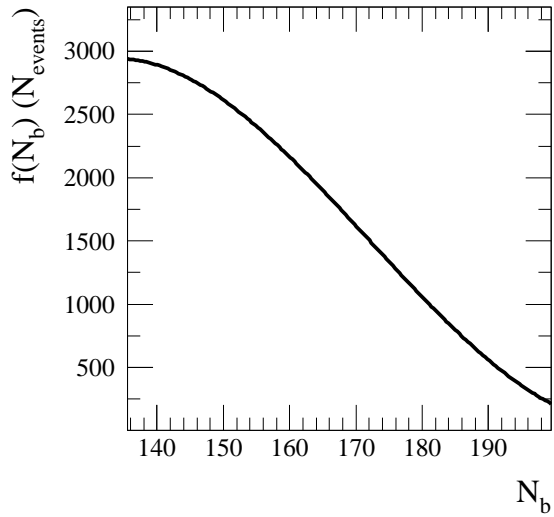}}
\hfill
\parbox{0.48\hsize}{
                    \caption{Baryon number of the Centauro fireball
                     produced in $\sqrt{s}=5.5~\times A$~TeV
                     $Pb+Pb$ collisions with impact parameter $0 < b < 5$~fm.}
                    \label{fig:Nb-distr}}
\end{figure}

The baryon number distribution of the Centauro fireball
produced in $\sqrt{s} = 5.5~\times A$~TeV $Pb+Pb$ collisions
is shown in Fig.~\ref{fig:Nb-distr}.

\paragraph*{Mass of the Centauro fireball.}
The Centauro fireball is a drop of deconfined quark matter
characterized by a temperature $T$ and a baryochemical potential $\mu_b$.
As the 
phenomenological model~\cite{Pana92,Pana94} predicts, it has a very
high baryochemical potential which does not permit the production of
$\bar u$ and $\bar d$ antiquarks. This state of the Centauro
fireball is unstable and after $\Delta t\sim 10^{-23}$~sec~\cite{Pana94}
gluons fragment into $s\bar s$ pairs and chemical equilibrium in the
fireball is achieved.In first-order perturbative QCD the energy density
of quark-gluon plasma containing $u$, $d$, $s$ quarks and gluons at
temperature $T$ close to the critical temperature $T_c$ is expressed
by (see e.g.~\cite{Muller92,Muller96} and references therein)
$$ \varepsilon = \varepsilon_g + \varepsilon_q + \varepsilon_s. $$
Here $q = u,d$. Gluon and quark contributions $\varepsilon_g$,
$\varepsilon_q$ and $\varepsilon_s$ are
$$ \begin{array}{rcl}
\varepsilon_g & = & \displaystyle \frac{8\pi^2}{15}T^4
  \left(1-\frac{15}{4\pi}\alpha_{\rm s} \right), \\[4mm]
\varepsilon_q & = & \displaystyle \frac{7\pi^2}{10}T^4
  \left(1-\frac{50}{21\pi}\alpha_{\rm s} \right)
 + \left(3 \mu_q^2 T^2 + \frac{3}{2\pi^2} \mu_q^4\right)
  \left(1-\frac{2}{\pi}\alpha_{\rm s} \right), \\[4mm]
\varepsilon_s & = & \displaystyle \gamma_s \left[
  \left(\frac{18T^4}{\pi^2}\right)
  \left(\frac{m_s}{T}\right)^2 K_2\left(\frac{m_s}{T}\right) +
 6\left(\frac{m_s T}{\pi}\right)^2
  \left(\frac{m_s}{T}\right) K_1\left(\frac{m_s}{T}\right)\right].
\end{array} $$
Here $K_i$ are $i$-order modified Bessel functions.
The strong coupling constant $\alpha_{\rm s}$ should be taken at a scale
$Q \approx 2\pi T$ and equals $\alpha_{\rm s} =0.3$ at a critical
temperature $T_c = 170$~MeV~\cite{Muller96}. The $\gamma_s$ is the
strangeness equilibration factor ($\gamma_s \approx 0.4$).
The net energy density for all degrees of freedom is given by
\begin{equation}
\varepsilon = \frac{37\pi^2}{30}T^4
 \left(1-\frac{110}{37\pi}\alpha_{\rm s} \right)
 + \left(3 \mu_q^2 T^2 + \frac{3}{2\pi^2} \mu_q^4 \right)
  \left(1-\frac{2}{\pi}\alpha_{\rm s} \right) + \varepsilon_s.
\label{eq:e_density}
\end{equation}
Here the quark chemical potential $\mu_q$ can be expressed via 
the baryochemical potential $\mu_b$ as $\mu_q = \mu_b / 3$.

\begin{figure}[ht]
\parbox{9cm}{\epsfbox{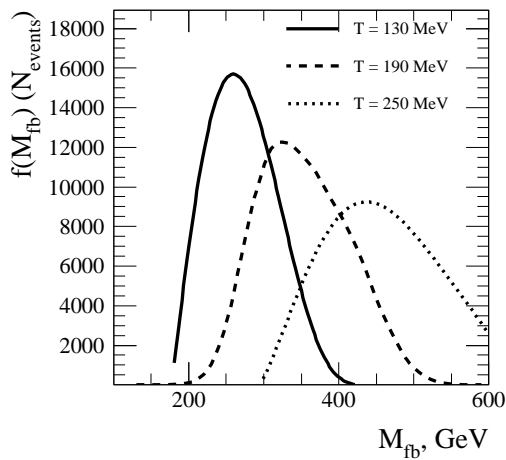}}
\hfill
\parbox{6.5cm}{
               \caption{Mass of the Centauro fireball produced in
                $\sqrt{s}=5.5~\times A$~TeV $Pb+Pb$ collisions
                with $\mu_b=1.8$~GeV and $T=130$, $190$ and $250$~MeV.}
               \label{fig:M-distr}}
\end{figure}

The other thermodynamical quantities of interest, pressure $P$ and
quark number density $n_q = N_q/V_{\rm fb}$ are obtained from equation
(\ref{eq:e_density}):
$$
P = \frac{1}{3}\varepsilon, \quad
n_q = \left( \frac{\partial P}{\partial \mu_q} \right)_T,
$$
\begin{equation}
n_q = 2 \left(\mu_q T^2 + \frac{\mu_q^3}{\pi^2}\right)
  \left(1-\frac{2}{\pi}\alpha_{\rm s} \right). \label{eq:q_density}
\end{equation}
Since the number of quarks $N_q$ in the primary Centauro fireball is
defined from the collision geometry as $N_q = 3N_b$ one can obtain
from (\ref{eq:Nb}) and (\ref{eq:q_density}) the volume of the
fireball $V_{\rm fb}$ to order ${\cal O}(\alpha_{\rm s})$:
\begin{equation}
V_{\rm fb} = \frac{3N_b}
    {\displaystyle 2 \left(\mu_q T^2 + \frac{\mu_q^3}{\pi^2}\right)}
  \left(1+\frac{2}{\pi}\alpha_{\rm s} \right).
        \label{eq:fbvol}
\end{equation}
When the volume of the fireball is defined one can easily obtain the
mass of the fireball from the energy density (\ref{eq:e_density}):
\begin{equation}
M_{\rm fb} = \varepsilon V_{\rm fb}.  \label{eq:fbmass}
\end{equation}
The distribution of the Centauro fireball mass produced in
$\sqrt{s} = 5.5~\times A$~TeV $Pb+Pb$ collisions with $\mu_b=1.8$~GeV
and $T=130$, $190$ and $250$~MeV is shown in Fig.~\ref{fig:M-distr}

\paragraph*{Kinematics of the Centauro fireball.}
Centauro events were observed in cosmic ray experiments in the very forward
region~\cite{Lattes1980,Cosmic}. We assume that the longitudinal
momentum distribution of the Centauro fireball obeys the same
scaling law of secondary particle production described by the empirical
formula established at lower energies for large $x_F$:
$$
dN/dx_F \sim (1-x_F)^n, \quad n \approx 3.
$$

Each constituent quark of the projectile nucleus which participates
in the formation of the fireball undergoes scattering by the
target nucleus. The transverse momentum distribution of a quark in the
fragmentation region can be expressed by the form
$$
dN_q/dp_T^2 \sim \exp\left(-\frac{p_T^2}{p_0^2}\right)
$$
with the slope $p_0 = 0.3$~GeV$/c$. Vector summation of the transverse
momenta of all interacting quarks gives the transverse momentum of
the produced Centauro fireball.

The rapidity range of the
fireball can be obtained from the following consideration.
The maximum rapidity of the fireball is reached when it carries the
full energy of the overlapping part of the projectile nucleus,
$E_{\rm max} = E_{\rm beam} N_b / A_{\rm beam}$:
$$
y_{\rm max} = \ln \frac{2E_{\rm max}}{M_{\rm fb}}.
$$
For example for central $Pb+Pb$ collisions at
$\sqrt{s} = 5.5~\times A$~TeV
and $N_b = 0.9\,A_{\rm beam} = 186$, assuming $T = 190$~MeV and
$\mu_b = 1.8$~GeV, one obtains the fireball mass
$M_{\rm fb} = 466$~GeV$/c^2$ and the maximum rapidity is
$$
y_{\rm max} = 7.69.
$$

However the nuclear stopping has to be considered as well,
as it is an important effect in heavy ion collisions.
It embodies the degree to which the energy of relative motion
of the two incident nuclei can be transferred into thermodynamical
degrees of freedom. The nuclear stopping can be expressed through
the rapidity shift $\Delta y_{\rm n.s.}$ of produced particles
compared to the maximum rapidity in $NN$ collisions.
The actual rapidity of the Centauro fireball is therefore defined
by the equation
\begin{equation}
  y_{\rm fb} = y_{\rm max} - \Delta y_{\rm fb}.
  \label{eq:y_fb}
\end{equation}
The value of $\Delta y_{\rm fb}$ is related to $\Delta y_{\rm n.s.}$
and is a crucial input parameter of the model on which the
observation of the Centauro events depends.
The average of the HIJING~\cite{Hijing} and VENUS~\cite{Venus}
predictions gives $\Delta y_{\rm n.s.} = 2.3$ but values in the range
$2.0 < \Delta y_{\rm n.s.} < 3.5$ can occur~\cite{Pana94}.

\paragraph*{Recoil system.}
Once the kinematics of the fireball are defined one can
calculate the momentum of the recoil system which consists of secondaries
from the target nucleus. Defining the 4-momentum of the Centauro fireball
as $p_{\scriptscriptstyle\rm Cn}$ and the 4-momentum of the recoil system
as $p_{\rm rec}$ we have the momentum conservation law as
$$ p_{\rm proj} + p_{\rm targ} = p_{\scriptscriptstyle\rm Cn} + p_{\rm rec}.$$
Let $\sqrt{s_{aa}}$ be the c.m.s.\ collision energy of the overlapping
fragments of the beam nuclei. If $\sqrt{s_{NN}}$ is the collision
energy per nucleon we obviously have $\sqrt{s_{aa}} = N_b \sqrt{s_{NN}}$
with $N_b$ defined by equation (\ref{eq:Nb}). Neglecting the mass of the
Centauro fireball in comparison with $\sqrt{s_{aa}}$ we obtain the mass
of the recoil system $M_{\rm rec}$ to be defined by the expression
$$
M_{\rm rec} = \sqrt{s_{aa}} (1 - \delta)^{1/2},
$$
where $\delta \approx 2 M_{\rm fb}\cosh(y_{\rm fb}) / \sqrt{s_{aa}}$.
For the rapidity of the recoil system $y_{\rm rec}$ the equation is as
follows:
$$
\sinh y_{\rm rec} \approx \frac{\delta/2}{(1-\delta)^{1/2}}.
$$
For values of the fireball rapidity shift $\Delta y_{\rm fb}$ of
several units, $\Delta y_{\rm fb} = 2 - 3$, one can conclude that the
recoil system carries almost the total energy $\sqrt{s_{aa}}$ of the
nuclear collision. In this approximation it is easy to show that
$\delta$ vanishes, and that hence the mass of the recoil system
$M_{\rm rec}$ is very close to the value of $\sqrt{s_{aa}}$ and
$y_{\rm rec}$ is small.

As an example, Table~\ref{tab:recoil-kin} gives the recoil mass and
rapidity in central $Pb+Pb$ collisions at $\sqrt{s} = 5.5~\times A$~TeV,
for $\sqrt{s_{aa}} = 1140$~TeV and when the Centauro fireball
mass is $M_{\rm fb} = 530$~GeV$/c^2$, for different values of
$\Delta y_{\rm fb}$.
\begin{table}[ht]
\centerline{
\begin{tabular}{|c|r@{.}lr@{.}lr@{.}lr@{.}l|} \hline
$\Delta y_{\rm fb}$  &  2&0      &  2&5      &  3&0      &  3&5 \\ \hline
$\displaystyle M_{\rm rec}/\sqrt{s_{aa}}$
                       &  0&93     &  0&96     &  0&97     &  0&98    \\
$y_{\rm rec}$          & $-0$&$07$ & $-0$&$04$ & $-0$&$03$ & $-0$&$02$
   \\ \hline
\end{tabular}}
\centerline{\parbox{0.9\hsize}{
\caption{Recoil system mass $M_{\rm rec}$ and rapidity $y_{\rm rec}$ in
$\sqrt{s}=5.5~\times A$~TeV $Pb+Pb$ collisions for
different values of the rapidity shift due to nuclear stopping
$\Delta y_{\rm fb}$} of the Centauro fireball.
  \label{tab:recoil-kin}}}
\end{table}
From this table it follows that the recoil system is produced in the
central rapidity region and, therefore, the secondary particles can be
detected by the central detector of any experiment. The composition
of the recoil system is unknown.

\begin{figure}[ht]
\parbox{9cm}{\epsfbox{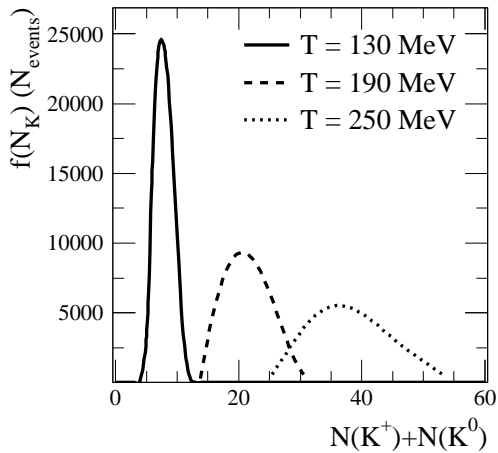}}
\hfill
\parbox{6.5cm}{
               \caption{Number of $K^+$ and $K^0$ emitted from the Centauro
                fireball produced in $\sqrt{s}=5.5~\times A$~TeV
                $Pb+Pb$ collisions with $\mu_b=1.8$~GeV and $T=130$,
                $190$ and $250$~MeV.}
               \label{fig:NK-distr}}
\end{figure}

\paragraph*{Strange quark matter fireball.}
As 
mentioned earlier gluons in the primary Centauro fireball fragment into
$s\bar s$ pairs and in this way chemical equilibrium is achieved.
The strange quark number density is given by the equation~\cite{Biro83}:
\begin{equation}
n_s = 1.37 \cdot 10^{-3}~\mbox{GeV}^3
      \left(\frac{T}{200~\mbox{MeV}}\right)
      K_2 \left(\frac{m_s}{T}\right),
  \label{eq:s_density}
\end{equation}
where $K_2(x)$ is a modified Bessel function of the second order.
Multiplied by the Centauro fireball volume $V_{\rm fb}$~(\ref{eq:fbvol})
equation~(\ref{eq:s_density}) gives the number of $s\bar s$ pairs
inside the fireball and, hence, the number of emitted $K$-mesons:
\begin{equation}
N_{\bar s} = N(K^+) + N(K^0) = n_s V_{\rm fb}.  \label{eq:nkaons}
\end{equation}
Fig.~\ref{fig:NK-distr} shows the distribution of the number of kaons
emitted from the Centauro fireball produced in
$\sqrt{s} = 5.5~\times A$~TeV $Pb+Pb$ collisions
with $\mu_b=1.8$~GeV and $T=130$, $190$ and $250$~MeV.
Before kaons are emitted from the fireball the total number of quarks
is $N'_q = 3N_b + 2N_{\bar s}$. Hence, the average energy per constituent
quark at this stage is
\begin{equation}
\epsilon'_q = \frac{M_{\rm fb}}{N'_q}. \label{eq:eq_prim}
\end{equation}
After $2 N_{\bar s}$ quarks have been emitted as kaons the mass of
the remaining SQM fireball is defined by the average quark
energy~(\ref{eq:eq_prim}) and the number of quarks in the fireball $N_q$:
$$
M'_{\rm fb} = N_q \epsilon'_q =
  M_{\rm fb} \left(1 - \frac{2N_{\bar s}}{N_q}\right).
$$
The emission of anti-strangeness is described as an isotropic decay of the
primary fireball into $N_{\bar s}$ kaons and the SQM fireball with the mass
$M'_{\rm fb}$.

\paragraph*{Decay of SQM fireball.}
After emission of kaons the primary Centauro fireball is transformed
into a slightly strange quark matter one which can have a long
life-time, of the order of $10^{-9}$~sec~\cite{Pana94}. At the final stage
of its evolution the SQM fireball decays into baryons and strangelets.
The latter are light droplets of strange quark matter with $A>6$,
high strangeness-per-baryon ratio $S/A \approx 1$ and small charge-to-mass
ratio $Z/A \approx 0$.
For simplicity, only one strangelet is formed in the SQM fireball
through random selection of $u$-, $d$- and $s$-quarks
among all the quarks of the fireball. If not the
complete strangeness content of the SQM fireball is transferred
to the strangelet, the remaining $s$-quarks form strange hyperons.
Baryons are formed in the fireball through random selection
of sets of three quarks among the quarks of the fireball.
Priority is given to the formation of nucleons and all quarks which
cannot be incorporated into nucleons produce strange hyperons.
The SQM fireball decays isotropically. We use the well-known event
generator {\sc Jetset}~\cite{Jetset} to perform further decays of kaons
and strange baryons.

\paragraph*{General characteristics of Centauro events.}
Table~\ref{tab:fb_chars} shows characteristics of Centauro events
in $Pb+Pb$ collisions at $\sqrt{s} = 5.5~\times A$~TeV. At given impact
parameter $b$, temperature $T$ and baryochemical potential $\mu_b$ we
calculate baryon number $N_b$, energy density $\varepsilon$, quark
number density $n_q$, volume of fireball $V_{\rm fb}$, mass of
primary fireball $M_{\rm fb}$, mass of SQM fireball $M'_{\rm fb}$,
strange quark number density $n_s$ and number of emitted kaons $N(K^{+,0})$.
\begin{table}[ht]
\centerline{
\begin{tabular}{|c|c|c||cccccccc|} \hline
$b$ & $\mu_b$ & $T$ & $N_b$ &
 $\varepsilon$ & $n_q$ & $V_{\rm fb}$ & $M_{\rm fb}$ & $M'_{\rm fb}$ &
   $n_s$ & $N(K^{+,0})$ \\
fm & GeV & MeV & & $\mbox{GeV/fm}^3$ & $\mbox{fm}^{-3}$ &
  $\mbox{fm}^3$ & GeV & GeV & $\mbox{fm}^{-3}$ & \\ \hline
0 & 1.8 & 130 & 186 &  4.3 &  6.7 &  83 & 357 & 344 & 0.13 & 11 \\
  &     & 190 & 186 &  7.7 &  9.2 &  61 & 466 & 423 & 0.48 & 28 \\
  &     & 250 & 186 & 13.6 & 12.5 &  45 & 607 & 515 & 1.14 & 50 \\ \cline{
  2-11}
  & 1.5 & 130 & 186 &  2.7 &  4.4 & 125 & 334 & 316 & 0.13 & 16 \\
  &     & 190 & 186 &  5.3 &  6.5 &  86 & 460 & 402 & 0.48 & 40 \\
  &     & 250 & 186 & 10.4 &  9.2 &  60 & 626 & 503 & 1.14 & 68 \\ \hline
5 & 1.8 & 130 & 114 &  4.3 &  6.7 &  51 & 219 & 212 & 0.13 &  6 \\
  &     & 190 & 114 &  7.7 &  9.2 &  37 & 286 & 260 & 0.48 & 17 \\
  &     & 250 & 114 & 13.6 & 12.5 &  27 & 372 & 315 & 1.14 & 31 \\ \hline
8 & 1.8 & 130 &  53 &  4.3 &  6.7 &  24 & 102 &  98 & 0.13 &  3 \\
  &     & 190 &  53 &  7.7 &  9.2 &  17 & 133 & 120 & 0.48 &  8 \\
  &     & 250 &  53 & 13.6 & 12.5 &  13 & 173 & 147 & 1.14 & 14 \\ \hline
\end{tabular}}
\centerline{\parbox{0.9\hsize}{
\caption{Properties of Centauro events for different fixed
values of impact parameter, temperature and baryochemical potential.}
\label{tab:fb_chars}}}
\end{table}
For some initial parameters of the model, especially for central
collisions ($b = 0$) and high temperature, Centauro events are
characterized by a high mass and a large number of kaons.
Nuclear collisions with large impact parameters could also produce
Centauro-type events but these events are characterized by
a smaller strange component.

The Centauro events 
observed in cosmic ray experiments are characterized by total,
or almost total, absence of the photonic component among secondary
particles. Since our model is based on the assumption that the
primary Centauro fireball consists of $u$ and $d$ quarks and gluons,
leading to the suppression of production of $\bar u$ and $\bar d$
antiquarks as described previously, it follows that the bulk of 
secondary hadrons seen in cosmic ray Centauros are baryons. 

\begin{figure}[htp]
\parbox{0.48\hsize}{\epsfxsize=\hsize \epsfbox{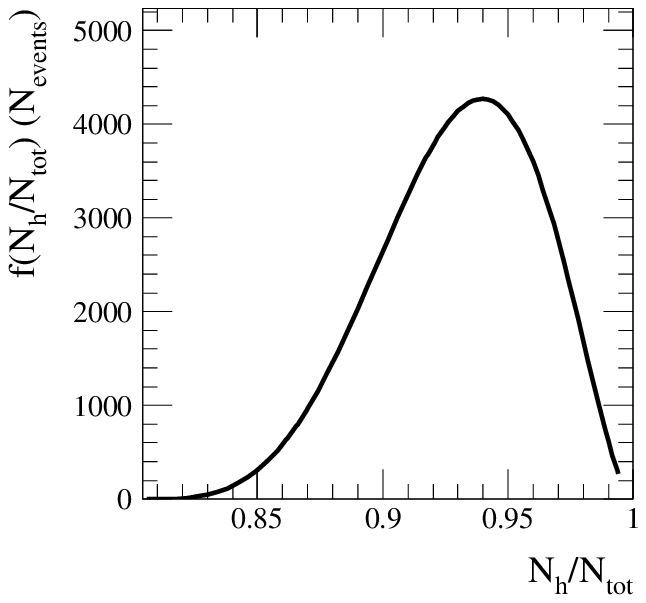}
                    \caption{Ratio of hadron to total multiplicities
                     (hadrons + photons) in Centauro events with
                     $\mu_b=1.8$~GeV and $T=190$~MeV produced in
                     $\sqrt{s}=5.5~\times A$~TeV $Pb+Pb$ collisions.}
                    \label{fig:N-ratio}}
\hfill
\parbox{0.48\hsize}{\epsfxsize=\hsize \epsfbox{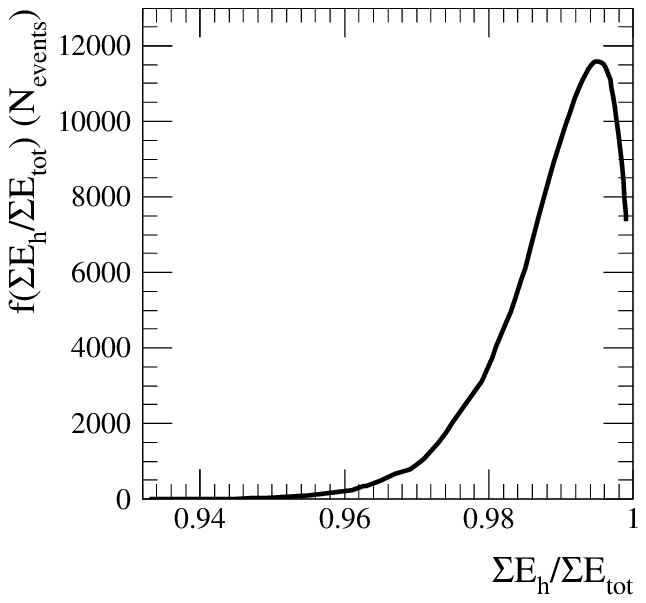}
                    \caption{Ratio of summed hadronic to total energies
                     in Centauro events with $\mu_b=1.8$~GeV and
                     $T=190$~MeV produced in
                     $\sqrt{s}=5.5~\times A$~TeV $Pb+Pb$ collisions.}
                    \label{fig:E-ratio}}
\end{figure}

Kaons
emitted from the primary fireball can decay into pions
which provide, in turn, an additional source of photons.
Overall however the
electromagnetic component of such an event is greatly suppressed.
Fig.~\ref{fig:N-ratio} shows the ratio of the hadron multiplicity to
the total multiplicity (hadrons + photons) in Centauro events with 
$\mu_b=1.8$~GeV and $T=190$~MeV produced in $\sqrt{s} = 5.5~\times A$~TeV
$Pb+Pb$ collisions. The ratio is very close to 1, with
mean value $\langle N_h / N_{\rm tot}\rangle = 0.93$, and the deviation of
this value from 1 is caused by the electromagnetic particles.
Fig.~\ref{fig:E-ratio} shows the ratio of the summed energy
of hadrons to the total energy in the same events. This ratio is
also very close to 1, with average value
$\langle \sum E_h / \sum E_{\rm tot}\rangle = 0.99$.
It should be noted that these ratios depend also on the
thermodynamical characteristics of the Centauro fireball, e.g.\ the higher
the temperature, the more kaons and therefore the more photons, are produced
and these ratios deviate further from 1.

\begin{figure}[htp]
\parbox[t]{0.48\hsize}{\epsfxsize=\hsize \epsfbox{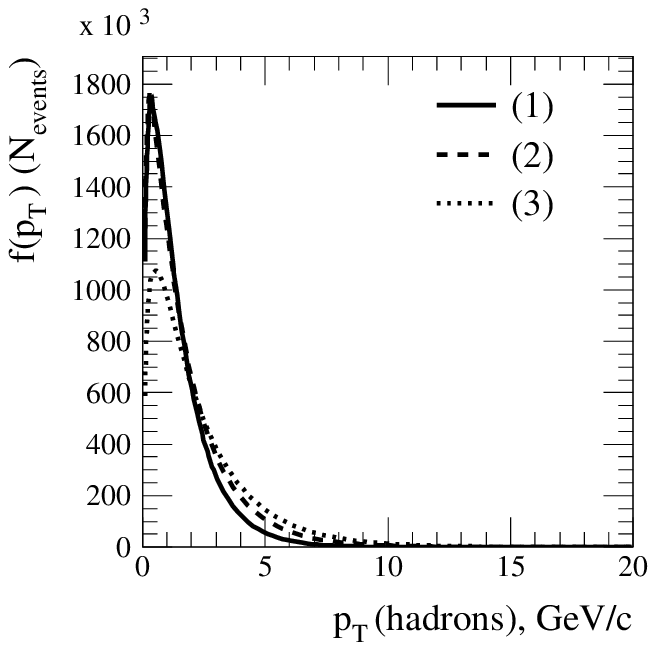}
                       \caption{Transverse momentum distribution of hadrons
                        in Centauro events
                        produced in $\sqrt{s}=5.5~\times A$~TeV
                        $Pb+Pb$ collisions
                        with $\mu_b=1.8$~GeV, $T=130$~MeV (1); $\mu_b=1.8$~GeV,
                        $T=190$~MeV (2) and $\mu_b=3.0$~GeV, $T=250$~MeV (3).}
                       \label{fig:pt-distr}}
\hfill
\parbox[t]{0.48\hsize}{\epsfxsize=\hsize \epsfbox{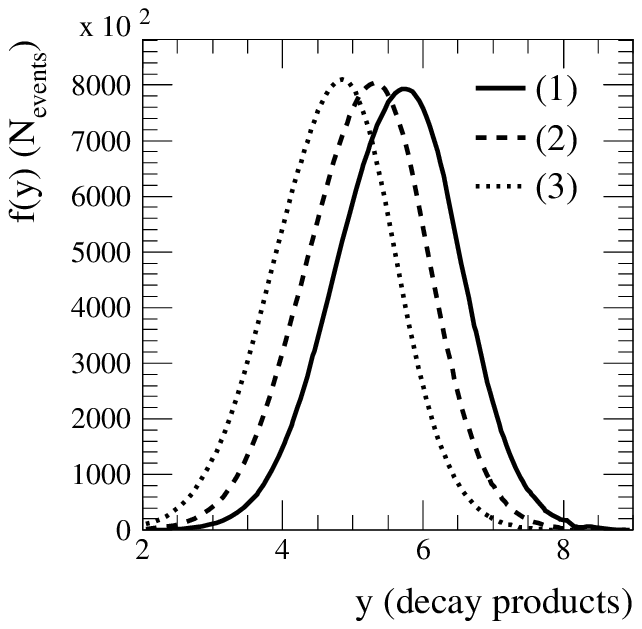}
                       \caption{Rapidity distribution of hadrons in Centauro
                        events
                        produced in $\sqrt{s}=5.5~\times A$~TeV
                        $Pb+Pb$ collisions
                        for three values of $\Delta y_{\rm fb} = 2.0$ (1),
                        $2.5$ (2) and $3.0$ (3).}
                       \label{fig:y-distr}}
\end{figure}

Secondary particles in
events arising from Centauro fireball decay have a larger
average transverse momentum in comparison with ordinary
hadronic interactions. The mean $p_T$ observed in cosmic
rays~\cite{Lattes1980,Cosmic} is $\langle p_T \rangle = 1.75$~GeV$/c$.
Fig.~\ref{fig:pt-distr} shows the transverse momentum distribution of
hadrons in Centauro events
produced in $\sqrt{s} = 5.5~\times A$~TeV $Pb+Pb$ collisions.
for three sets of baryochemical potential $\mu_b$ and temperature $T$:
$\mu_b=1.8$~GeV, $T=190$~MeV; $\mu_b=1.8$~GeV, $T=250$~MeV and $\mu_b=3.0$~GeV,
$T=250$~MeV. The average $p_T$ in such events is $\langle p_T \rangle
= 1.34$~GeV$/c$, $1.47$~GeV$/c$ and $1.75$~GeV/$c$ respectively.
Conventional hadronic events, as predicted by HIJING,
have average transverse momentum $\langle p_T \rangle = 0.44$~GeV$/c$
which is $2-4$ times smaller than 
in Centauro events.

The rapidity distribution of decay products of the Centauro fireball
clearly depends on the nuclear stopping power. In Fig.~\ref{fig:y-distr}
the rapidity distributions of secondary particles are shown for three
values of the fireball rapidity shift $\Delta y_{\rm fb} = 2.0$, $2.5$
and $3.0$. Obviously all secondary particles from the Centauro fireball decay
are distributed in the very forward region, as 
observed in cosmic rays.
However, the complete event within which the Centauro fireball
is produced contains also the particles of the recoil system which
fall in the central region, $y \approx 0$, according to
Table~\ref{tab:recoil-kin}.

\begin{figure}[htp]
\parbox[t]{0.48\hsize}{\epsfxsize=\hsize \epsfbox{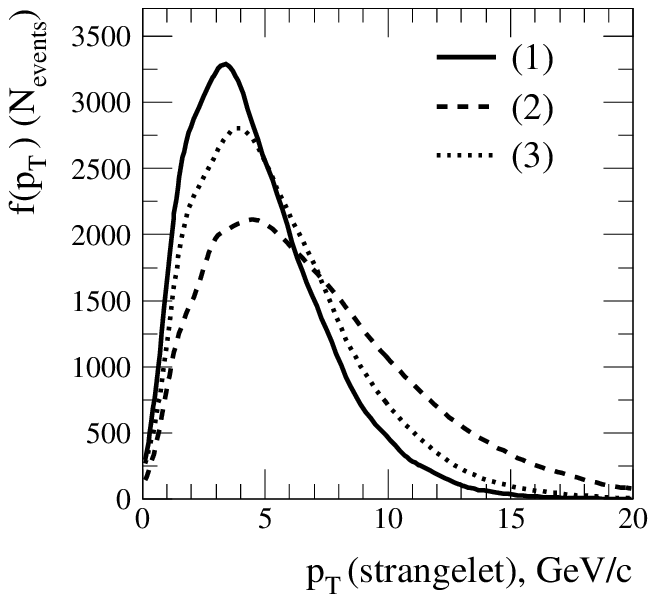}
                       \caption{Transverse momentum distribution of
                        strangelets 
                        formed in the decay of Centauro fireballs
                        produced in $\sqrt{s}=5.5~\times A$~TeV
                        $Pb+Pb$ collisions for
                        $\mu_b=1.8$~GeV, $T=130$~MeV (1); $\mu_b=1.8$~GeV,
                        $T=190$~MeV (2) and $\mu_b=3.0$~GeV, $T=250$~MeV (3).}
                       \label{fig:pt-distr-str}}
\hfill
\parbox[t]{0.48\hsize}{\epsfxsize=\hsize \epsfbox{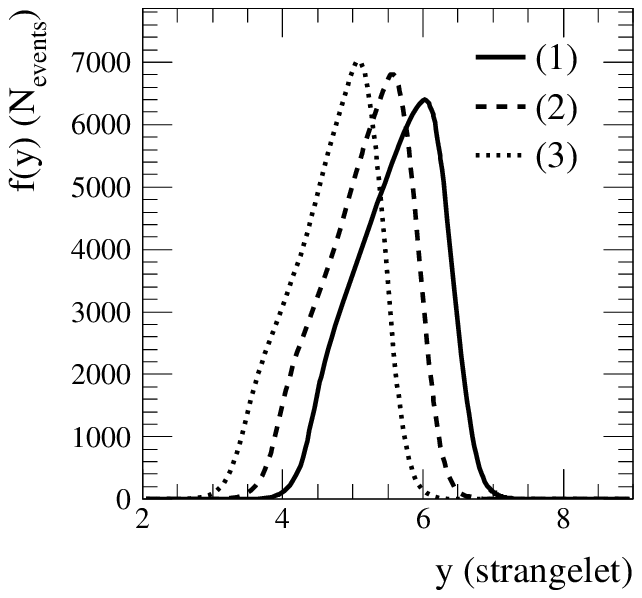}
                       \caption{Rapidity distribution of strangelets
                        formed in the decay of Centauro fireballs
                        produced in $\sqrt{s}=5.5~\times A$~TeV
                        $Pb+Pb$ collisions for
                        $\Delta y_{\rm fb} = 2.0$ (1), $2.5$ (2) and
                        $3.0$ (3).}
                       \label{fig:y-distr-str}}
\end{figure}

The kinematics of the strangelets which may be produced in the Centauro
events are similar to the hadron kinematics. Due to their larger
mass, the strangelets have larger transverse momentum.
Fig.\ref{fig:pt-distr-str} shows the transverse momentum distribution of
strangelets formed in the decay of Centauro fireballs produced in 
$\sqrt{s} = 5.5~\times A$~TeV $Pb+Pb$ collisions for three sets of the
thermodynamical parameters, $\mu_b=1.8$~GeV, $T=190$~MeV;
$\mu_b=1.8$~GeV, $T=250$~MeV and $\mu_b=3.0$~GeV, $T=250$~MeV.
The rapidity distribution of the strangelets is shown in
Fig.\ref{fig:y-distr-str} for three values of the fireball rapidity
shift $\Delta y_{\rm fb} = 2.0$, $2.5$ and $3.0$.

%

\section{Detection of Centauro events with CASTOR}

In this section we present simple calculations of the geometrical
detection efficiency of the CASTOR detector
at the LHC~\cite{Aris1,Aris2,Aris3}. The detector will probe the
very forward, baryon rich region in $\sqrt{s} = 5.5~\times A$~TeV
$Pb+Pb$ collisions.
It will be installed at $\sim 16.4$~m from the interaction point,
will be azimuthally symmetric around the beam pipe and as close to it
as possible. The detector's inner radius will be $R_{\rm in} = 2.8$~cm
and its outer radius $R_{\rm out} = 15$~cm, enabling it to cover
pseudorapidities $5.6 < \eta < 7.0$.
Here we present Monte Carlo calculations of the charged particle and photon
multiplicities and electromagnetic and hadronic energies reaching the front
face of the detector.

\begin{figure}[ht]
\parbox{0.48\hsize}{\epsfxsize=\hsize \epsfbox{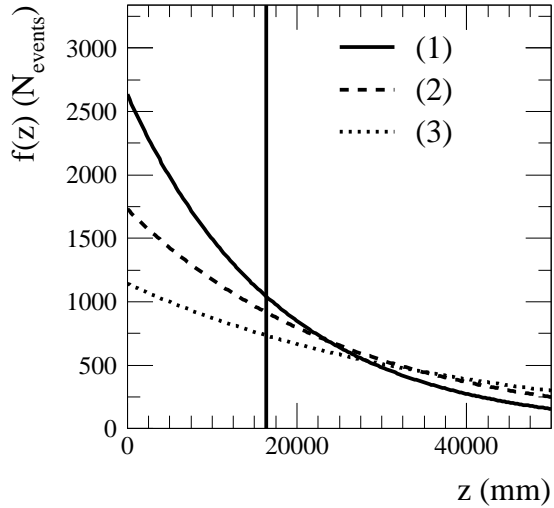}}
\hfill
\parbox{0.48\hsize}{
                    \caption{Decay length of the SQM fireball at
                      $T=250$~MeV, $\mu_b = 1.8$~GeV and $\Delta
                      y_{\rm fb} = 3.0, 2.5, 2.0$. The vertical line
                      at $z=16.4$~m shows the detector positions.}
                    \label{fig:z-path}}
\end{figure}

Due to a large lifetime in the rest frame, $\tau\sim 10^{-9} $~sec and
a high rapidity, the SQM fireball can decay far apart from the beam
interaction point. The decay length of the fireball before its decay
depends on the fireball mass and the rapidity shift $\Delta y_{\rm
  fb}$ (\ref{eq:y_fb}). In the Fig.~\ref{fig:z-path} the path length
distribution of the fireball is shown for $T=250$~MeV, $\mu_b =
1.8$~GeV and three values of the rapidity shift $\Delta y_{\rm fb} =
3.0, 2.5$ and $2.0$.
The detector positions is shown by the vertical solid line. As it is
seen, the significant number of SQM fireballs decays beyond the
detector, i.e. their decay products will not be detected. Only
$K$-mesons which emitted from the primary Centauro fireball at the
initial stage of its evolution can be detected.

\begin{figure}
\parbox[t]{0.48\hsize}{\epsfxsize=\hsize \epsfbox{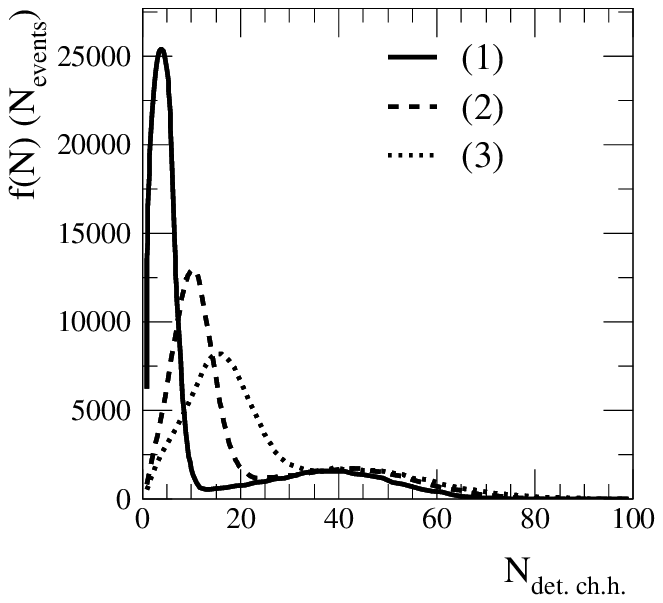}
                       \caption{Charged hadron multiplicity in the detector
                        for Centauro events produced in
                        $\sqrt{s}=5.5~\times A$~TeV $Pb+Pb$ collisions
                        for fixed $\Delta y_{\rm fb} = 2.5$,
                        $\mu_b=1.8$~GeV and
                        $T=130$ (1), $190$ (2) and $250$~MeV (3).} 
                       \label{fig:Nh-det-T}}
\hfill
\parbox[t]{0.48\hsize}{\epsfxsize=\hsize \epsfbox{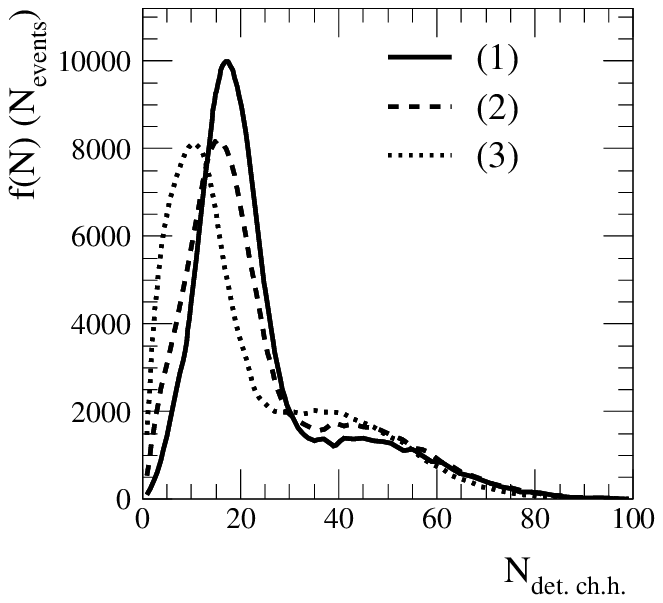}
                       \caption{Charged hadron multiplicity in the detector
                        for Centauro events produced in
                        $\sqrt{s}=5.5~\times A$~TeV $Pb+Pb$ collisions
                        for fixed $\mu_b=1.8$~GeV, $T=250$ and 
                        $\Delta y_{\rm fb} = 2.0$ (1), $2.5$ (2) and $3.0$ (3).
}
    \label{fig:Nh-det-Dy}}
\end{figure}

The detection of secondary particles of the Centauro events also
strongly depends on the parameters of the model, namely the
thermodynamical variables $\mu_b$ and $T$ which influence the mass of
the fireball and the fireball rapidity shift due to nuclear stopping
$\Delta y_{\rm fb}$. We compare the ability of CASTOR to detect
Centauro fireballs produced at different $T$ and fixed $\mu_b$ and
$\Delta y_{\rm fb}$, as well as at different $\Delta y_{\rm fb}$ and
fixed thermodynamical parameters.  Fig.~\ref{fig:Nh-det-T} shows the
charged hadron multiplicity distribution at the detector for fixed
$\Delta y_{\rm fb} = 2.5$ and $\mu_b=1.8$~GeV and three values of the
fireball temperature $T=130$, $190$ and $250$~MeV. The dependence of
the charged hadron multiplicity on the rapidity shift for fixed
$\mu_b=1.8$~GeV and $T=250$~MeV is shown in Fig.~\ref{fig:Nh-det-Dy}:
the curves correspond to $\Delta y_{\rm fb} = 2.0$, $2.5$ and
$3.0$. The shape of these distributions is affected by the decay
length distribution shown in Fig.~\ref{fig:z-path}. The sharp peaks at
small multiplicity corresponds to the events when the SQM fireball
decays behind the detector, and only $K$-mesons contribute to the
detected multiplicity.

The number of detected charged hadrons for each set of parameters 
has to be compared to the total number of charged hadrons generated
in the Centauro event. The geometrical detection efficiency of
the detector for different values of the model parameters is tabulated in
Table~\ref{tab:eff-ch-had}. The fourth column gives the average
geometrical efficiency of charged hadron detection $e_{\rm ch.had.}$ in
terms of $\mu_b$, $T$ and $\Delta y_{\rm fb}$.
\begin{table}[ht]
\centerline{
\begin{tabular}{|ccc|cc|} \hline
$\mu_b$, GeV & $T$, MeV &
 $\Delta y_{\rm fb}$
 & $e_{\rm ch.had.}$ &
                                                $e_{\rm str.}$ \\
    \hline
1.8        & 130      & 2.5                   & 0.38 & 0.67 \\
1.8        & 190      & 2.5                   & 0.42 & 0.57 \\
1.8        & 250      & 2.5                   & 0.46 & 0.46 \\[2mm]
1.8        & 250      & 2.0                   & 0.42 & 0.61 \\
1.8        & 250      & 3.0                   & 0.35 & 0.36 \\
3.0        & 250      & 2.5                   & 0.39 & 0.45 \\
    \hline
\end{tabular}}
\centerline{\parbox{0.8\hsize}{
\caption{Geometrical detection efficiency $e$ of charged hadrons and
strangelets in terms of $\mu_b$, $T$ and $\Delta y_{\rm fb}$.}
\label{tab:eff-ch-had}}}
\end{table}
The charged hadron multiplicity of Centauro-type events in the detector
is rather small, not more than 120 with mean values $30-60$.
The photon multiplicity is much smaller as can be seen from
Fig.~\ref{fig:N-ratio}.

\begin{figure}[htb]
\parbox{0.48\hsize}{\epsfxsize=\hsize \epsfbox{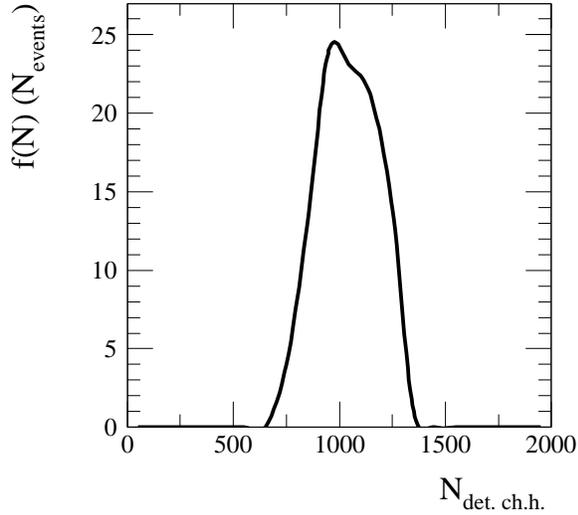}}
\hfill
\parbox{0.48\hsize}{
\caption{Charged hadron multiplicity in the detector in 
         conventional $\sqrt{s}=5.5~\times A$~TeV $Pb+Pb$
         collisions as predicted by HIJING.} 
\label{fig:Nh-det-hij}}
\end{figure}

The small multiplicity of the Centauro events is in sharp
contrast to what is expected for conventional nuclear collisions
where the multiplicity is calculated to be of the order of
a few thousand. Fig.~\ref{fig:Nh-det-hij} shows the charged hadron
multiplicity detected by CASTOR in conventional
$\sqrt{s} = 5.5~\times A$~TeV $Pb+Pb$ collisions with impact parameters
$0 < b < 5$~fm as predicted by HIJING.
The multiplicity in the conventional events is several times
higher than
that in the Centauro-type events, with an average detected
multiplicity of 1000 which is about 20 times that in
Centauro events.

\begin{figure}[ht]
\parbox{0.48\hsize}{\epsfxsize=\hsize \epsfbox{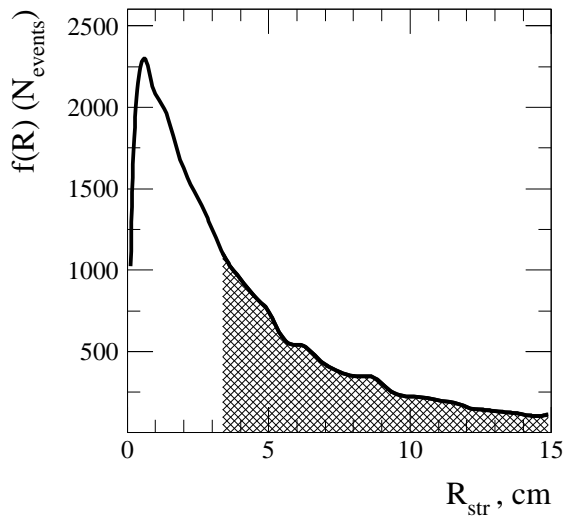}}
\hfill
\parbox{0.48\hsize}{
\caption{Distribution of hit radius on the detection plane
         of strangelets formed in the decay of Centauro
         fireballs produced in $\sqrt{s}=5.5~\times A$~TeV
         $Pb+Pb$ collisions with $\mu_b=1.8$~GeV, $T=250$~MeV
         and $\Delta y_{\rm fb} = 2.5$.
         Shaded area corresponds to the detector surface.}
       \label{fig:R-strlet}}
\end{figure}

Because of their larger mass strangelets are boosted forward more
than ordinary hadrons are. Therefore strangelets have a tendency to fly
closer to the beam. The distribution of the radius of strangelet hits on
the detection plane for strangelets formed in the decay of Centauro
fireballs produced in $\sqrt{s}=5.5~\times A$~TeV $Pb+Pb$ collisions with
$\mu_b=1.8$~GeV, $T=250$~MeV and $\Delta y_{\rm fb} = 2.5$ is shown in
Fig.~\ref{fig:R-strlet}.
The shaded area corresponds to the detector surface.
The geometrical efficiency of strangelet detection $\epsilon_{\rm str.}$
for different values of the model parameters is given in the last
column of Table~\ref{tab:eff-ch-had}.

Results obtained for somewhat different detector configurations
and model parameters have been presented in~\cite{Ewa2001,Jaipur1997}.

\section*{Conclusions}

We presented quantitative simulated results on the
observation of Centauro events in heavy ion collisions at LHC energies.
The phenomenological model of Centauro events originally introduced
in~\cite{Pana92,Pana94} gives a transparent explanation of such events.
On the basis of this model we constructed the quantitative model and the
event generator {\sc CNGEN} which provides a tool to estimate the
geometrical detection efficiency of Centauro events and associated
strangelets.

The possibility to observe Centauro events depends strongly on the
parameters of the model such as thermodynamical characteristics and the
nuclear stopping power.

The signatures for observation of Centauro events with the CASTOR
detector can be summarized as follows:
\begin{itemize}
\item small detected charged particle multiplicity,
      $\langle N_{\rm ch.~h.}\rangle = 50$ compared to
      $\langle N_{\rm ch.~h.} \rangle = 1000$ in
      conventional hadronic events;
\item significant predominance of the detected hadron multiplicity
      which can be characterized by the average hadron-to-all particles ratio 
      $\langle N_h / N_{\rm tot}\rangle > 0.9$ while this ratio is equal to
      $\langle N_h / N_{\rm tot}\rangle = 0.6$ in 
      conventional hadronic events;
\item significant predominance of the detected hadron energy deposited in the
      detector, $\langle \sum E_h / \sum E_{\rm tot} \rangle = 0.99$
      compared to
      $\langle \sum E_h / \sum E_{\rm tot} \rangle = 0.80$ in 
      conventional hadronic events;
\item large average transverse momentum, $\langle p_T \rangle > 1$~GeV$/c$
      compared to $\langle p_T \rangle = 0.44$~GeV$/c$ in
      conventional hadronic events.
\end{itemize}

The model affords the possibility of strangelet production in the
decay of the Centauro fireball. If such objects exist and behave according
to the model, for $A>6$ their geometrical detection efficiency
in CASTOR would be about $40-60\%$. The geometrical detection
efficiency of the other secondaries of the Centauro events is
also satisfactory.

\vspace*{5mm} 
This work was partly supported by Polish State Committee
for Scientific Research grant \hbox{No. 2P03B 011 18} and 
\hbox{SPUB-M/CERN/P-03/DZ 327/2000}.  The authors would like to thank
Zbigniew W{\l}odarczyk for useful remarks.

\vspace*{5mm} 
The source code of the event generator {\sc CNGEN} can be obtained by
request from \verb|kharlov@mx.ihep.su|.

\end{document}